\documentstyle[preprint,revtex]{aps}
\begin{document}
\draft
\preprint{}
\begin{title} Giant Shapiro resonances in a flux driven \\
Josephson junction necklace \end{title}
\author{G. Chu and Jorge V. Jos{\'e}}
\begin{instit}         Department of Physics,
        Northeastern University,
        Boston, MA 02115
\end{instit}
\thispagestyle{empty}
\begin{abstract}
We present a detailed study of the dynamic response of a ring of $N$ equally
spaced Josephson junctions to a time-periodic external flux,
including screening current effects. The dynamics are described by the
resistively shunted Josephson junction model, appropriate for proximity
effect junctions, and we include Faraday's law for the flux. We find that the
time-averaged $I-V$ characteristics  show novel
{\em subharmonic giant Shapiro voltage resonances}, which  strongly depend
on having phase slips or not, on $N$, on the inductance and on the external
drive frequency. We include an estimate of the
possible experimental parameters needed to observe these quantized voltage
spikes.

\end{abstract}
\newpage
\narrowtext
Motivated by significant advances in microfabrication techniques of
Josephson junction ($JJ$) arrays,
many contemporary studies have focused on understanding the dynamic
response of two-dimensional $JJ$ arrays [1-7]. One type of array
with well-defined uniform properties is made of
superconducting-normal-superconducting ($SNS$) proximity effect junctions.
Several novel collective locked-in states have been discovered which are
manifested in the current versus voltage ($IvsV$) characteristics as {\rm
plateaus}. These giant Shapiro steps are a macroscopic manifestation
of diverse underlying collective vortex oscillating states [1-5]. Most recent
studies, experimental and theoretical [1-5,7], have considered  current driven
two-dimensional $JJ$ arrays. In this paper we study  the dynamic response of
a ring of $N$ equally spaced $JJ$, referred to as a {\em Josephson Junction
Necklace} ($JJN$), driven by a perpendicular external time-dependent flux
$\Phi _e(t)$, including self-induced magnetic field effects.
The motivation of this study is two-fold; first, to understand
the dynamic response of a JJ array to a $\Phi _e(t)$, and
second, to assess the importance of geometry in this
response. As we show in this paper, the physics of the JJN is, in many respects
qualitatively and quantitatively different from that of a current driven
planar array.

The $JJN$ is shown schematically in the inset of Fig. 1.
This model is a natural extension of the standard $rf$ $SQUID$ to
$N$ junctions  instead of one. The external time-dependent flux considered here
produces a Faraday voltage $V_e(t)$ of the form, $-\frac{d}{dt}\Phi
_e(t)=V_e(t)=V_{dc}+V_{ac}\cos2\pi \tilde {\nu} t.$
We also include the self-induced magnetic flux $\Phi _i$ due to the screening
currents, of importance in a proximity effect $JJ$.
We model the dynamics of the order-parameter phase of
the $JJN$ in terms of the resistively shunted Josephson junction model ($RSJ$)
defined by the current
\begin{equation}
I(t)=I_c\sin [\phi _{i}-\phi _{i+1}- \psi_{i,i+1}]+
\frac{\hbar}{2eR}\frac{d}{dt}
[\phi _{i}-\phi _{i+1}- \psi_{i,i+1}].
\end{equation}
In this equation $\Delta _{\mu}\phi= \phi _{i}-\phi _{i+1}$ is the
phase difference for the $\mu$-th junction
denoted as $\times$ in the inset of Fig. 1. $I_c$ is its critical current and
$R$ its shunt resistance, both
assumed to be the same for all junctions in the necklace. Here
$\displaystyle\psi _{i,i+1}(t)=\frac{2\pi}{\Phi _0}\int _i^{i+1}
\vec{A}\cdot d\vec{\ell}=\frac{2\pi}{N}\frac{\Phi}{\Phi _0}
\equiv \frac{2\pi}{N} f(t)$, where $\Phi (t)$ is the {\it total} flux
through the ring of radius $r$,
$f(t)=({{\Phi _e(t)+\Phi _i(t)})/ {\Phi _0}}= f_e+f_i$,
and $\Phi_0=\displaystyle\frac{h}{2e}$ is the quantum of flux.
The induced flux is given by $\Phi _i= L I(t)$, where $L$ is the geometrical
inductance of the ring. It is convenient to write Eq(1) in dimensionless units
with $ i(\tau)=I(t)/I_c$, $ \tau=t (2eRI_c/\hbar) $, $ v=V/(RI_c)$, $\ell= L
\displaystyle \frac{2\pi I_c }{\Phi_0}$, and
$\nu _{\phi}=\displaystyle\frac{\hbar \tilde {\nu}}{2eRI_c}$.
In these units Eq.(1) becomes
\begin{equation}
i(\tau)= \sin [\frac {2\pi}{N}\biggr ([f(\tau)]_k - f(\tau)\biggr )]+
\frac {2\pi}{N}\frac{d}{d\tau}
\biggr ( [f(\tau)]_k - f(\tau) \biggr ).
\end{equation}
In writing this equation we have used the fact that the current is conserved
in the ring which, together with the periodic boundary conditions (p.b.c.)
$\phi _{N+1}=\phi _1$, implies that the phase difference $\Delta \phi =
\frac {2\pi}{N} [f(\tau)]_k$, where $[f]_k$ stands for the nearest integer
function defined as $[f(\tau )]_k={\rm int}(f+\displaystyle\frac{1}{2})$.
The specific form of the applied external flux considered in this paper
is $f_e={\Phi _e/\Phi_0}=-(-f_e^0+ v_{dc}/{2\pi}\tau+
\frac{v_{ac}}{(2\pi)^2 \nu _{\phi}} \sin 2\pi \nu _{\phi} \tau)$, with
$f_e^0\equiv f_e(\tau=0)$. Substituting this expression for $f_e$ in Eq(2) we
get
\begin{eqnarray}
& &\frac{\ell}{N} \frac{d}{d\tau}i(\tau)+ \left\{ i(\tau)-
\sin[\frac {2\pi}{N} [f(\tau)]_k + \frac{1}{N}(-2\pi f_e^0+
v_{dc}\tau+\frac{v_{ac}}{2\pi \nu _{\phi}}\sin 2\pi \nu _{\phi}\tau
-{\ell}\, i(\tau))] \right\} \nonumber\\
&=&\frac{1}{N}(v_{dc}+v_{ac}\cos 2\pi \nu _{\phi}\tau)+
\frac {2\pi}{N} \frac{d}{d\tau}[f(\tau)]_k. \end{eqnarray}
Eq(3) is the general equation describing the response of the JJN to
a time-dependent external flux,
and understanding the structure of its time-dependent solutions is the central
goal of this paper. To find $i(\tau )$ from Eq(3) we need to know
$[f(\tau)]_k$. As shown below, we can use the ground state
properties  ($f_e=f_e^0$) of the $JJN$  to obtain solutions for
$[f(\tau)]_k$ in the relevant physical regimes. We start then
by considering the ground state properties of the JJN including inductive
effects, which  appear not to have been discussed  before for
$N\geq 2$ \cite{charge}.
{}From the p.b.c. and current conservation it follows that the phase difference
for each and all junctions is $\Delta _{\mu}\phi={(2\pi k)/N}$, with
$k=0,1,2,3,..,N-1$. When the self-inductance is neglected,
the ground state energy per junction $E_g/N$, in normalized units, is
$\epsilon _N= \frac{2\pi E_g}{\Phi _0I_cN}=
-cos[\frac{2\pi}{N}([f_e^0]_k-f_e^0)]$,
with the corresponding normalized ground state current $i_N$ obtained from
$i_N=-\frac{\partial \epsilon _N}{ \partial f_e^0}=\sin[\frac{2\pi}{N}
([f_e^0]_k-f_e^0)]$. For a periodic JJN we note the important  symmetry of the
ground state energy $f_e^0\rightarrow f_e^0+1$ and $f_e^0\rightarrow f_e^0-1$,
which allows the analysis to be restricted to $f_e^0\epsilon [0,1]$.
In  Fig. 1 we show $\epsilon _N(\ell =0)$
(solid  line) for $N=10$. The difference between curves $I$ and $II$
is that in $II$ there is a discontinuity in $\epsilon _N$ at $f_e^0=1/2$,
whereas curve $I$ is continuous. Physically, as
$f_e^0$ increases from zero to $f_e^0=(1/2)^{-}$ the JJN can either follow
curve $I$ or curve $II$. Following $II$ entails absorbing a
quantum of flux ($k=1$) while following $I$ implies no change in the flux
trapped by the JJN. Of course, in equilibrium the system will prefer to follow
$II$ rather than $I$, which is a higher energy state. However,
if there is an  external $f_e(\tau )$ present the JJN will be allowed to follow
curve $I$ as well.
We shall call the process following curve $II$ a {\it phase slip} ($PS$) case
while the one that follows $I$ the {\it no phase slip} ($NPS$) process.
For the $\ell \neq 0$ case, the ground state energy becomes
$\epsilon _N=-cos[\frac{2\pi}{N}([f]_k-f)]+\frac {2\pi ^2f_i^2}{\ell N}$
and the corresponding normalized current is
$i_N=\sin[\frac {2\pi}{N} ([f]_k-f)]$. From these
equations one finds that the total flux in the ring is given by
$f=f_e^0+\frac{\ell}{2\pi}\sin[\frac{2\pi}{N}([f]_k-f)]$,
which is a self-consistent transcendental equation that, for a
given set of values for $f_e^0$ and $\ell$, can be solved numerically. We
note that in contrast to the $N=1$ case, where there is a critical value
$\ell _c=1$ that separates
non-hysteretic ($\ell > 1$) from hysteretic ($\ell <1$) behavior,
the $JJN$ with $N\geq 2$ is {\it always} hysteretic. The size of the
hysteresis loop grows as $\ell$ increases from zero until
$\ell\geq N$, when the hysteresis loop covers the whole $f_e^0\epsilon
[0,1]$ range. The metastability of the $JJN$ is evident from
the history-dependent ground state energy shown as a dashed line
in Fig. 1. In the $PS$ case we see that there are two values of $f_e^0$ for
which $\epsilon _N$ has a discontinuity, one while ramping up $f_e^0$
the other when decreasing it. The specific $f_e^0$ values at which the
discontinuities take place depend on $\ell$ and $N$. The $NPS$ case
is, again, represented by continuous processes.

With the above information we are now ready to discuss the dynamic properties
of the $JJN$. We start by introducing the important
parameter  $\kappa ^2\equiv \nu _{_{\Phi}}/\nu _{_{\phi }}$, with
$\nu _{\Phi}$ the characteristic
frequency for the relaxation of magnetic fields. In the linearized regime
of the Josephson term in Eq(3), $\nu _{\Phi}= R/L$,
and   $\kappa=\sqrt{{\Phi_0}\over{2\pi I_c L}}\equiv \sqrt{1/{\ell}}$.
This parameter measures the importance of screening effects:
when $\kappa =\infty$ they are negligible while for $\kappa \leq 1$
they are all important. Thus
we want to study the dynamics of the JJN as a function of $\kappa$ in
the two extreme regimes of $PS$ and $NPS$.
For simplicity we shall write $\ell$ instead of $\kappa$ in
our results, although the connection between $\kappa=\sqrt{1/{\ell}}$
is only valid in the linearized regime. Most of the time the value of
$[f(\tau )]_k$ is constant except when $f(\tau)$ passes through the
phase slip points of Fig. 1. In this case $[f(\tau )]_k$ is discontinuous and
its derivative $\frac{d}{d\tau}[f(\tau)]_k$ has a $\delta$-function character.
We use this information in evaluating the time-averaged current given in Eq(3).
We start by considering the simplest case when $\ell =0$, i.e. when there
is no screening. In the $NPS$ case $<i_N>$ can be calculated analytically
since  $\frac{d}{d\tau}[f(\tau)]_k=0$,  and the integral
$<i_N> =\lim_{\tau\to\infty}\frac{1}{\tau}\int
_{-\frac{\tau}{2}}^{\frac{\tau}{2}}
ds \sin [\frac{1}{N}(2\pi k -2\pi f_e^0+v_{dc}s +\frac{v_{ac}}{2\pi
\nu _{\phi}}\sin 2\pi \nu _{\phi}s)] +v_{dc}/N$,  gives $<i_N>=v_{dc}/N$, when
$v_{dc}\ne mN\nu _{\phi}$, and
$ <i_N>=\sin [\frac{2\pi}{N}(k-f_e^0)] J_{-m}(\frac{v_{ac}}{2\pi N\nu _{\phi}})
+v_{dc}/N$, whenever $v_{dc}=mN\nu _{\phi}$, with $m=0,1,2,\cdots$.
In deriving this result we have used the identity $e^{iz\sin\phi}=
\sum_{-\infty}^{\infty}J_{-m}(z) e^{-im\phi}$, with
$J_{-m}(z)$ the Bessel function of integer  order.
Note that when $v_{dc}=mN\nu _{\phi}$ the initial condition
$f_e^0$ plays a crucial role since $<i_N>=<i_N>(f_e^0)$, and thus
when $v_{dc}\equiv mN\nu _{\phi}$ there are finite intervals of $<i_N>$ for
each value of $v_{dc}$. These {\it giant Shapiro resonances} ($GSR$) are shown
in Fig. 2(a) for $N=10$, $\nu _{\phi}=0.01$ and $v_{ac}=1$. We note that
the size of the current intervals for which there are resonances are comparable
to one. The averaged current in the $PS$ can be evaluated analytically
following a similar logic as in ref. \cite{tom}. The
qualitative result is that there are {\it subharmonic} resonances
for $v_{dc}=\frac {m}{n}{N\nu _{\phi}}$, with
$n=1,2\cdots$. The analytic analysis
does not give, however, the order of magnitude of these {\it subharmonic}
resonances. To resolve this question we evaluated $<i_N>$ in the $PS$ case
numerically. The results for $N=10, \nu _{\phi}=.01, v_{ac}=1$ are shown in
Fig. 2(c). These results were obtained as follows:
first the time interval was divided in $\Delta \tau$ segments, with
$\nu _{\phi}\Delta \tau=10^{-4}$, and the  $v_{dc}$ interval into 24
pieces. Next, for each value of $v_{dc}$ we took 100 initial conditions for
$f_e^0$. Whenever $f_e(\tau)$ reaches a discontinuity point, which is
determined with a precision of $10^{-6}$, we changed the value of
$[f(\tau)]_k$.
Note that the scale in Fig. 2(c) is about an order of magnitude smaller
than in Fig. 2(a).

We now consider the general case with $\ell \neq 0$. To evaluate the
time-averaged current in this case we need to solve the differential
equation given in Eq(3).
This is not a simple ordinary differential equation
since there is an implicit transcendental dependence of $i_N(\tau)$ through
the sine function. The equation is solved by using a fourth order Runge-Kutta
method following a similar procedure as in the $\ell =0$ case,
for parameter values similar to those of Fig. 2(c) but with
$\ell =50.0$, or $\kappa =0.1414$. For the $NPS$ case the results are shown in
Fig. 2(b). There we see  {\it subharmonic resonances} but of
magnitude larger than in Fig. 2(c). We call these
resonances then {\it giant subharmonic Shapiro resonances} ($GSSR$).
In the $PS$ case and for the same parameter values as in Fig. 2(b),
Fig. 2(d) shows, in contrast,  that the subharmonic resonances remain of the
same order of magnitude as in the $\ell =0$ limit.
We also calculated  the spectral function, not shown here,
defined as $S(2\pi \nu _{\phi}) = \lim_{\tau\rightarrow\infty}{\vert
{1\over \tau}{\int_0^\tau i_N(s)
e^{i2\pi \nu _{\phi}s}ds}\vert^2}$. It is found that at a resonance the
current as a function of time is a periodic function but with a complicated
subharmonic structure.

The results presented in Fig. 2 show that the $JJN$ is capable of exhibiting
truly $GSSR$ in the $NPS$ case although they are also present, but of a smaller
magnitude, in the $PS$ case. To further understand the properties
of these resonances we calculated the magnitude of the resonance spikes
widths $\Delta <i_N>$ as a function
of $\nu _{\phi}, \sqrt{\ell} =1/{\kappa} ,v_{ac}$ and $N$. The corresponding
results
are shown in Fig. 3(a-d) for the $1/2$ resonance, for both the $NPS$ case
($\diamond$) and the $PS$ case ($\times$). The significant characteristic of
Figs. 3(a-c) is that for the $NPS$ case there are maximum or optimal
$\Delta <i_N>$ values
as a function $\nu _{\phi}, \sqrt {\ell}$ and $N$. The situation is less
clear for the $PS$ case although small maxima can be seen in Figs. 3(a-c).
These specific maximum values are listed in the figure caption.
Note that as a function of ${\ell}$, and as seen in
Figs. 2,  $\Delta <i_N>$ is larger for
the $NPS$ case in the  large $\ell$ limit but it is slightly smaller in
the $PS$ case in the small $\ell$ regime. A similar suppression of the
 $1/2$ step width for large $\nu _{\phi}$ was found in [3,7] for a current
driven square array. The suppression was not found, however, for integer
steps. In contrast, we found that the $n=1$ step width as a function of
$\nu _{\phi}$ behaves in a qualitatively similar way as in the $1/2$ case.
In Fig. 3(d) we show the behavior
of  $\Delta <i_N>$ as a function of $v_{ac}$. Contrary to what is seen in one
JJ or in the 2D-array calculations, $\Delta <i_N>$ grows monotonically up
to about $v_{ac}=3$ and then it appears to become periodic for the $NPS$ case
but is still aperiodic for the $PS$ limit. We checked that these results
are stable against an increase in the number of periods used
to calculate the averages.
All the results shown in Fig. 3 indicate that the qualitative and quantitative
properties of the $JJN$ are inherently different from those obtained previously
for one JJ or for 2D-arrays.

These results allow us to make a rough estimate of the appropriate experimental
parameters for which these resonances can be seen experimentally
taking the typical values from experiments \cite{expt,cinci}.
For a lattice constant $a\sim 10 \mu m$ and with separation between junctions
of $b\sim  2 \mu m$, with the optimal value of $N=16$, we get a diameter of
$d\sim 61 \mu m$. The  critical currents are between $1\times 10^{-2}$ to
$10$ $mA$, so that using the estimate for the inductance for a ring \cite{indu}
of $L=1.25 \mu _0d$, one gets the values of $\sqrt {\ell}$ between
$1.73\sim 54.84$ which are within the range of maximum values shown in
Fig. 3(b).

In conclusion, we have presented a detailed analysis of the response of a
Josephson junction necklace to an external time-dependent flux. The main
result of our analysis is that there are giant subharmonic Shapiro
resonances, mainly in the no phase slip or ``fast" regime.
Furthermore, we found that these
resonances have a strong dependence on the number of junctions $N$,
normalized frequency $\nu _{\phi}$ and effective inductance $\ell$.
An estimate was also provided for the possible experimental conditions under
which these resonances may be seen.

\acknowledgments
We thank D. Dom\' \i nguez for useful discussions and suggestions.
This work was supported in part by $NSF$ grant DMR-9211339, the  Pittsburgh
supercomputing center under grant PHY88081P, the Donors of the Petroleum
Research Fund, under grant ACS-PRF\#22036-AC6.
and the Northeastern University Research and Development Fund.

\figure { Normalized Ground state energy for $N=10$ as a function of $f_e^0$.
Solid line $\ell =0$ and dashed line $\ell =5.0$.
The curves $I$ and $II$ represent the no-phase-slip ($NPS$) and phase-slip
($PS$) processes, respectively. The inset gives a schematic representation
of the $JJN$ model studied here.}
\figure { The time-averaged normalized current $<i_N>$ vs $v_{dc}$ for $N=10$,
$\nu _{\phi}=0.01$ and $v_{ac}=1.0$. (a) the $NPS$ and (c) $PS$ cases, both
with $\ell =0$. (b) the $NPS$ and (d)  $PS$ cases with $\ell =50.0$.
The results were obtained as described in the text.
The $\displaystyle\frac{1}{2}$ and $\displaystyle\frac{1}{3}$
subharmonic resonances are indicated by arrows.}

\figure{ Renormalized current width $\Delta <i_N>$ for the 1/2-resonance  as
a function of (a) $\nu _{\phi}$, (b) $ \sqrt{\ell}$, (c) $N$
and (d) $v_{ac}$ in the $NPS$ ($\diamond$) and $PS$ ($\times$) cases.
In (a) $v_{ac}=1, N=10$ and $\ell =50.0$. In (b) $\nu _{\phi}=0.01$, $N=10$ and
$v_{ac}=1$. In (c) $ v_{ac}=1,\ell =50.0$ and $\nu _{\phi}=.01$. In
(d) $ \nu _{\phi}=0.01,\ell =50.0$ and $N=10$. The approximate maxima in Figs.
3(a-d)
are located at $\nu _{\phi}^{max}\sim 0.0251$, $\sqrt{{\ell}^{max}}
\sim 5.012$ and $N^{max}\sim 16$, respectively.}

\end{document}